\documentclass[twocolumn,showpacs,amsmath,amssymb,prd]{revtex4}

\usepackage{graphicx}
\usepackage{dcolumn}
\usepackage{bm}
\usepackage{epsf}
\allowdisplaybreaks

\begin{document}

\title{Improved numerical stability of stationary black hole
evolution calculations}

\author{Hwei-Jang Yo$^{1,2}$, Thomas W. Baumgarte$^{3,1}$,
        and Stuart L. Shapiro$^{1,4}$}

\affiliation{$^1$Department of Physics,
	University of Illinois at Urbana-Champaign, Urbana, Illinois 61801}

\affiliation{$^2$Institute of Astronomy and Astrophysics, 
	Academia Sinica, Taipei 115, Taiwan, Republic of China}

\affiliation{$^3$Department of Physics and Astronomy, Bowdoin College, 
	Brunswick, Maine 04011}

\affiliation{$^4$Department of Astronomy \& NCSA, 
	University of Illinois at Urbana-Champaign, Urbana, Illinois 61801}

\begin{abstract}
We experiment with modifications of the BSSN form of the Einstein field
equations (a reformulation of the ADM equations) and demonstrate how these
modifications affect the stability of numerical black hole evolution
calculations.  We use excision to evolve both non-rotating and rotating 
Kerr-Schild black holes in octant and equatorial symmetry, and without any
symmetry assumptions, and obtain accurate and stable simulations for
specific angular momenta $J/M$ of up to about $0.9 M$.
\end{abstract}
\maketitle

\section{Introduction}
\label{intro}

Binary black holes are among the most promising sources for the
gravitational wave laser interferometers currently under development,
including LIGO, VIRGO, GEO, TAMA and LISA.  The identification and
interpretation of possible signals requires theoretically predicted
gravitational wave templates.  For the late epoch of the binary
inspiral, numerical relativity is the most promising tool for 
the computation of such templates.  

The numerical simulation even of single black holes has encountered
numerous difficulties, which presumably arise from the
complexity of Einstein's equations, the existence of a singularity
inside the black hole, and the gauge (or coordinate) freedom inherent
in general relativity.  Some recent developments, however, have led
to significant and very promising advances.

Traditionally, most numerical relativity simulations were based on the
$3+1$ decomposition of Arnowitt, Deser and Misner (ADM, \cite{adm62}),
which has been shown to develop instabilities often \cite{bonc2,baut}.
To avoid these instabilities, a number of hyperbolic
formulations of Einstein's equations have been developed (see, e.g.,
\cite{bonc2,ay99} as well as \cite{r98} and references therein).
Alternatively, Shibata and Nakamura \cite{shim} and Baumgarte and
Shapiro \cite{baut} introduced a modification of the original ADM
equations that involves a conformal-traceless decomposition and the
introduction of a new auxiliary variable (see Section \ref{Bf} below).
This formulation, now commonly referred to as the BSSN formulation, has
led to significant improvements over the original ADM equations, and
has been widely adopted (e.g.~\cite{baut2,shim2,shim3,shim4,alcm4,alcm5}).

Singularities inside black holes were traditionally avoided by using
``singularity avoiding'' coordinate conditions, including maximal or
polar slicing \cite{smal,eard,barj,plst8586}.  Typically, these conditions lead
to grid pathologies that cause codes to crash after relatively short
times.  An alternative strategy takes advantage of the fact that the
black hole exterior is causally disconnected from the interior, so
that a region of the black hole interior including the singularity can
be excised from the computational grid.  These ``singularity
excision'' techniques \cite{unrw84,thoj2,seie,thoj3,annp,schm2} have led to
large improvements in the simulation of single black holes, and are a
promising tool for binary black hole evolutions (see \cite{bras} for
preliminary results).

The application of singularity excision requires a coordinate system
that is regular across black hole horizons, allowing smooth horizon
penetration.  For single black holes, one such coordinate system is
the Kerr-Schild form, which can be used to represent both
Schwarzschild and Kerr black holes (and which is also invariant under
boosts).  Binary black hole initial data based on Kerr-Schild
coordinates can be constructed by solving the constraint equations of
general relativity for ``corrections'' arising from superpositions of
two boosted Kerr-Schild black holes
\cite{mrhs99,bisn,marp,stp02,pct02}.

Alcubierre and Br\"ugmann \cite{ambb01} recently combined the BSSN
formalism with a particularly simple singularity excision method to
evolve single black holes in Kerr-Schild coordinates.  Restricting the
evolution to octant symmetry, all fields settle down to equilibrium
(of the finite-difference equations), and no instabilities are
encountered.  If, however, the symmetry assumption is relaxed,
instabilities develop and the code crashes after a few hundred $M$.
Similar findings were reported in \cite{s00}, where a completely
independent formulation and implementation was adopted.  Improvements
over these results were discussed in \cite{klst01,lpsd02,ls02}, but
even these do not completely eliminate the instabilities for
evolutions without symmetry assumptions.

Recent results suggest that adding constraints to the evolution
equations affects the numerical stability of the system
(e.g.~\cite{klst01,d87,ys01,kllpsst01,ygsh02}; see \cite{kwb02} for an
illustration in electrodynamics).  In this paper we experiment with
adding the new constraints that appear in the BSSN formulation to the
evolution equation of the new auxiliary functions, and, following
\cite{ygsh02}, the Hamiltonian constraint to the evolution equation
for the spatial metric.  We also experiment with schemes for imposing
algebraic constraints on the conformally related metric and extrinsic
curvature, as well as with different shapes for the excised region
inside the black hole.

With these modifications we obtain evolutions of single black holes
that last over several thousand $M$, independent of any symmetry
assumptions, without encountering any evidence of a growing
instability.  These findings hold both for static black holes
and for rotating black holes with specific angular momentum of 
up to $0.9M$.

The paper is organized as follows: We summarize the BSSN formulation
in Sec.~\ref{Bf} and black hole spacetimes in Kerr-Schild coordinates
in Section \ref{idfiksbh}.  Our modifications of the BSSN scheme are
described in Sec.~\ref{mar}.  In Section \ref{nr} we present results
of our simulations for both static and rotating BHs.  We summarize and
discuss the implications of our findings in Sec.~\ref{conc}.  We also
include an Appendix that explains our evaluation of the ADM mass and
angular momentum.  Throughout the paper we adopt geometrized units
with $G=c=1$.

\section{The BSSN formulation}
\label{Bf}

We write the metric in the ADM form
\begin{equation} \label{adm_metric}
ds^2 = - \alpha^2 dt^2 + \gamma_{ij} (dx^i + \beta^i dt)(dx^j + \beta^j dt),
\end{equation}
where $\alpha$ is the lapse function, $\beta^i$ is the shift vector,
and $\gamma_{ij}$ is the spatial metric.  Throughout this paper, Latin
indices are spatial indices and run from 1 to 3, whereas Greek indices
are space-time indices and run from 0 to 3.

The Einstein equations can then be decomposed into the Hamiltonian
constraint ${\mathcal H}$ and the momentum constraints ${\mathcal
M}_i$
\begin{eqnarray}
   {\mathcal H} &\equiv& R - K_{ij}K^{ij} + K^2 = 0, \label{ham1} \\
   {\mathcal M}_i&\equiv& D_j K^{j}_{~i} - D_i K = 0, \label{mom1}
\end{eqnarray}
and the evolution equations 
\begin{eqnarray} 
\frac{d}{dt} \gamma_{ij} & = & - 2 \alpha K_{ij},\label{gdot1} \\
\frac{d}{dt} K_{ij} & = & - D_i D_j \alpha + \alpha ( R_{ij}
        - 2 K_{i\ell} K^\ell_{~j} + K K_{ij}).       \label{Kdot1}
\end{eqnarray}
Here we have assumed vacuum $T_{\alpha\beta} = 0$ and have used
\begin{equation}
\frac{d}{dt} = \frac{\partial}{\partial t} - {\mathcal L}_{\beta},
\end{equation}
where ${\mathcal L}_{\beta}$ is the Lie derivative with respect
to $\beta^i$.   $D_i$ is the covariant derivative associated with
$\gamma_{ij}$, $R_{ij}$ is the three-dimensional Ricci tensor
\begin{eqnarray} \label{ricci}
   R_{ij} &=& \frac{1}{2} \gamma^{k\ell} \left( \gamma_{kj,i\ell}
    +\gamma_{i\ell,kj}-\gamma_{k\ell,ij}-\gamma_{ij,k\ell}\right)
    \nonumber\\ & & + \gamma^{k\ell} \left( \Gamma^m{}_{i\ell}
    \Gamma_{mkj} - \Gamma^m{}_{ij} \Gamma_{mk\ell} \right),
\end{eqnarray}
and $R$ is its trace $R = \gamma^{ij} R_{ij}$.  

In the BSSN formalism \cite{shim,baut}, the above ADM equations are
rewritten by introducing the conformally related metric $\tilde
\gamma_{ij}$
\begin{equation}\label{confgij}
\tilde \gamma_{ij} = e^{- 4 \phi} \gamma_{ij},
\end{equation}
with the conformal factor chosen so that the determinant $\tilde \gamma$ 
of $\tilde \gamma_{ij}$ is unity
\begin{equation}
e^{4 \phi} = \gamma^{1/3}.
\end{equation}
The traceless part of the extrinsic curvature $K_{ij}$, defined by
\begin{equation}
A_{ij} = K_{ij} - \frac{1}{3} \gamma_{ij} K,
\end{equation}
where $K = \gamma^{ij} K_{ij}$ is the trace of the extrinsic
curvature, is conformally decomposed according to
\begin{equation}
\tilde A_{ij} = e^{- 4 \phi} A_{ij}.
\end{equation}
The conformal connection functions $\tilde{\Gamma}^i$, initially defined as
\begin{equation} \label{Gamma}
\tilde{\Gamma}^i \equiv \tilde{\gamma}^{jk} \tilde{\Gamma}^i_{jk} 
	= - \tilde{\gamma}^{ij}_{~~,j},
\end{equation}
are regarded as independent variables in this formulation.

The evolution equations of BSSN formulation can be written as 
\begin{eqnarray}
\frac{d}{dt}\phi & = & 
	-\frac{1}{6}\alpha K,\label{eq:evolphi}\\
\frac{d}{dt}\tilde{\gamma}_{ij} & = &
	-2\alpha\tilde{A}_{ij},\label{eq:evolg}\\
\frac{d}{dt}K & = &
	\alpha\left(\tilde{A}_{ij}\tilde{A}^{ij}+\frac{1}{3}K^2
	\right)-\gamma^{ij}D_iD_j\alpha \nonumber \label{eq:evolK}\\
\frac{d}{dt}\tilde{A}_{ij} & = &
	\alpha\left(K\tilde{A}_{ij}-2\tilde{A}_{ik} 
   	\tilde{A}^k{}_j\right)\nonumber\\ 
   	&&+e^{-4\phi}\big(\alpha R_{ij} -D_iD_j\alpha\big)^{TF},
   	\label{eq:evolA}\\
\partial_t\tilde{\Gamma}^i & = & 2\alpha\left(\tilde{\Gamma}^i_{jk}
    \tilde A^{jk}-\frac{2}{3}\tilde{\gamma}^{ij}K_{,j}
    +6\tilde{A}^{ij}\phi_{,j}\right) \nonumber\\
   &&-2\tilde{A}^{ij}\alpha_{,j}+\tilde{\gamma}^{jk}\beta^i{}_{,jk}+
	\frac{1}{3}
    \tilde{\gamma}^{ij}\beta^k{}_{,jk}+\beta^j\tilde{\Gamma}^i{}_{,j}
    \nonumber\\
   &&-\tilde{\Gamma}^j\beta^i{}_{,j} 
    +\frac{2}{3} \tilde{\Gamma}^i\beta^j{}_{,j}.\label{eq:evolGamma2}
\end{eqnarray}
Here the superscript $TF$ denotes the trace-free part of a tensor.
The Ricci tensor $R_{ij}$ can be written as a sum of two pieces
\begin{equation}
   R_{ij} = \tilde{R}_{ij} + R^{\phi}_{ij},
\end{equation}
where $R^{\phi}_{ij}$ is given by
\begin{eqnarray}
   R^{\phi}_{ij}&=&-2\tilde{D}_i\tilde{D}_j\phi-2\tilde{\gamma}_{ij}
    \tilde{D}^k\tilde{D}_k\phi\nonumber\\
   &&+4\tilde{D}_i\phi\tilde{D}_j\phi-4\tilde{\gamma}_{ij}\tilde{D}^l\phi
   \tilde{D}_l\phi,
\end{eqnarray}
while, with the help of the $\tilde \Gamma^i$, $\tilde{R}_{ij}$ 
can be expressed as
\begin{eqnarray}
   \tilde{R}_{ij}&=&-\frac{1}{2}\tilde{\gamma}^{mn}\tilde{\gamma}_{ij,mn}
     +\tilde{\gamma}_{k(i}\tilde{\Gamma}^k{}_{,j)}
     +\tilde{\Gamma}^k\tilde{\Gamma}_{(ij)k} \nonumber \\
   &&+\tilde{\gamma}^{mn}\left(2\tilde{\Gamma}^k{}_{m(i}\tilde{\Gamma}_{j)kn}
     +\tilde{\Gamma}^k{}_{in}\tilde{\Gamma}_{kmj}\right).
\end{eqnarray}
The new variables are tensor densities, so that their Lie 
derivatives are
\begin{eqnarray}
   {\mathcal L}_\beta\phi&=&\beta^k\phi_{,k} + \frac{1}{6}\beta^k{}_{,k}, \\
   {\mathcal L}_\beta\tilde{\gamma}_{ij}&=&\beta^k\tilde{\gamma}_{ij,k}
    +2\tilde{\gamma}_{k(i}\beta^k{}_{,j)}-\frac{2}{3}\tilde{\gamma}_{ij}
    \beta^k{}_{,k},\\
   {\mathcal L}_\beta K&=& \beta^kK_{,k},\\
   {\mathcal L}_\beta\tilde{A}_{ij}&=&\beta^k\tilde{A}_{ij,k}
    +2\tilde{A}_{k(i}\beta^k{}_{,j)}-\frac{2}{3}\tilde{A}_{ij} \beta^k{}_{,k}.
\end{eqnarray}
The Hamiltonian and momentum constraints (\ref{ham1}) and (\ref{mom1})
can be rewritten as
\begin{eqnarray}
{\mathcal H} & = & e^{-4\phi}\left(\tilde{R}-8\tilde{D}^i\tilde{D}_i\phi
	-8\tilde{D}^i\phi\tilde{D}_i\phi\right)
   \label{ham2}\nonumber\\
   &&\qquad\qquad\qquad
   +\frac{2}{3}K^2-\tilde{A}_{ij}\tilde{A}^{ij} =0\\
{\mathcal M}^i&=&\tilde{D}_j\tilde{A}^{ij}+6\tilde{A}^{ij}\phi_{,j}-
	\frac{2}{3} \tilde{\gamma}^{ij}K_{,j}=0, \label{mom2}
\end{eqnarray}
where $\tilde{R}=\tilde{\gamma}^{ij}\tilde{R}_{ij}$.

\section{Black Holes in Kerr-Schild coordinates}
\label{idfiksbh}

The ingoing Kerr-Schild form of the Kerr metric is given by 
\cite{mrhs99,chas92}
\begin{equation} \label{ksf1}
   ds^2=(\eta_{\mu\nu} + 2H\ell_\mu\ell_\nu)dx^\mu dx^\nu,
\end{equation}
where $\eta_{\mu\nu}={\rm diag}(-1,1,1,1)$ is the Minkowski metric in
Cartesian coordinates, and $H$ a scalar function.  The vector
$\ell_\mu$ is null both with respect to $\eta_{\mu\nu}$ and
$g_{\mu\nu}$,
\begin{equation}
\eta^{\mu\nu}\ell_\mu\ell_\nu=g^{\mu\nu}\ell_\mu\ell_\nu=0,
\end{equation}
and we have $\ell^2_t=\ell^i\ell_i$.  The general Kerr-Schild BH
metric has
\begin{equation}
        H = \frac{Mr}{r^{2} + a^{2}\cos^{2} \theta} \label{3}
\end{equation}
and
\begin{equation}
        \ell_{\mu} = \left(1, \frac{rx + ay}{r^{2} + a^{2}}, \frac{ry -
        ax}{r^{2} + a^{2}}, \frac{z}{r}\right).  \label{4}
\end{equation}
Here $M$ is the mass of the Kerr BH, $a = J/M$ is the specific angular
momentum of the BH, and $r$ and $\theta$ 
are auxiliary spheroidal coordinates defined in
terms of the Cartesian coordinates by
\begin{equation}
   \frac{x^{2} + y^{2}}{r^{2} + a^{2}} + \frac{z^{2}}{r^{2}} = 1 \label{5}
\end{equation}
and $z = r \cos \theta$.  The event horizon of the BH is located at
\begin{equation}
r_{\rm eh} = M+\sqrt{M^2-a^2}.\label{reh}
\end{equation}

Comparing (\ref{ksf1}) with the ADM metric (\ref{adm_metric}) one
identifies the lapse function $\alpha$, shift vector $\beta_i$ and the
spatial 3-metric $\gamma_{ij}$ as 
\begin{eqnarray} \label{ksf2}
              \alpha &=& (1+2H)^{-1/2}, \\
             \beta_i &=& 2H\ell_i,\\
         \gamma_{ij} &=& \eta_{ij} + 2H\ell_i\ell_j.
\end{eqnarray}
We can see here that these variables all extend smoothly through the horizon
and their gradients near the horizon are well-behaved..
Given these metric quantities, the extrinsic curvature $K_{ij}$ 
can be computed from (\ref{gdot1})
\begin{eqnarray}
   K_{ij}&=&2\alpha H\ell^k(\ell_i\ell_jH_{,k}
    +2H\ell_{(i|}\partial_k\ell_{|j)})\nonumber\\
    &&+2\alpha(\ell_{(i}\partial_{j)}H+H\partial_{(i}\ell_{j)}),\\
    K&=&2\alpha^3(1+H)\ell^iH_{,i}+2\alpha H\ell^i{}_{,i}.
\end{eqnarray}

In the static case $a=0$, the above expressions reduce to the
Schwarzschild expressions in in-going Eddington-Finkelstein form
\cite{edda58}
\begin{eqnarray}
      H &=& M/r,\nonumber\\
      \ell_\mu &=& (1,x_i/r),\nonumber\\
      K_{ij}&=&\frac{2M}{ r^4(1+2M/r)^{1/2}}\left[r^2\eta_{ij}
       -(2+\frac{M}{r})x_ix_j\right],\\
      K&=& \frac{2M}{r^2(1+2M/r)^{3/2}}(1+3M/r).\nonumber
\end{eqnarray}
where $M$ is the total mass-energy and $r^2 = x^2 + y^2 + z^2$. 

\section{Adjusting the BSSN equations}
\label{mar}

For a solution of the BSSN equations to be equivalent with a solution
of the ADM equations, the new auxiliary variables have to satisfy new
constraint equations.  In particular, $\tilde A_{ij}$ has to be
traceless
\begin{equation} \label{tra} 
{\mathcal A} \equiv  
	\tilde{\gamma}^{ij}\tilde{A}_{ij} = 0,
\end{equation}
the determinant of the conformally related metric $\tilde \gamma_{ij}$
has to be unity
\begin{equation} \label{deg}
{\mathcal D} \equiv \det(\tilde{\gamma}_{ij}) - 1 = 0,
\end{equation}
and the conformal connection functions $\tilde \Gamma^i$ have to
satisfy the identity (\ref{Gamma})
\begin{equation}
{\mathcal G}^i \equiv 
	\tilde{\Gamma}^i - \tilde{\gamma}^{jk}\tilde{\Gamma}^i{}_{jk}
	= 0 \label{gi}.
\end{equation}
These conditions can be viewed as new constraints, in addition to the
Hamiltonian and momentum constraints (\ref{ham2}) and (\ref{mom2}).

In an unconstrained evolution calculation, the constraints are
monitored only as a code check.  It may be advantageous, however, either to 
enforce at least some of the constraints during the evolution,
or to add evolution constraint equations to the evolution equations.

Alcubierre and Br\"ugmann \cite{ambb01}, for example, found 
improved stability properties when the algebraic constraint
(\ref{tra}) is enforced.  This can be achieved by replacing
$\tilde{A}_{ij}$ with
\begin{equation} \label{AB1}
\mbox{AB1:~~~~}   
\tilde{A}_{ij} \leftarrow \tilde{A}_{ij}-\frac{1}{3}\tilde{\gamma}_{ij}
   \tilde{\gamma}^{mn}\tilde{A}_{mn}
\end{equation}
after each time step.  They also found improvements, at least in octant
symmetry, when using the conformal connection function $\tilde
\Gamma^i$ only in places where it is differentiated, and instead use,
by virtue of (\ref{gi}), the contraction of the Christoffel symbols
everywhere else
\begin{equation} \label{AB2}
\mbox{AB2:~~~~}
\tilde{\Gamma}^i \leftarrow 
\tilde{\gamma}^{jk}\tilde{\Gamma}^i_{jk}.
\end{equation}

Other ways of imposing some of the constraints are possible
(compare \cite{lpsd02}).  While the rule AB1 (equation \ref{AB1})
is appealing in that it treats all components of $\tilde A_{ij}$
identically, we found particularly stable results by dynamically
evolving only five of the six components of $\tilde A_{ij}$, and
computing the $zz$ component using the algebraic constraint
(\ref{tra})
\begin{equation}
   \tilde{A}_{zz} = -\frac{\tilde{A}_x{}^x+\tilde{A}_y{}^y+\tilde{A}_{xz}
   \tilde{\gamma}^{xz}+\tilde{A}_{yz}\tilde{\gamma}^{yz} }{
   \tilde{\gamma}^{zz}}.\label{azz}
\end{equation}
We similarly determine $\tilde \gamma_{zz}$ from the other five
metric components using the algebraic constraint (\ref{deg})
\begin{eqnarray}
   \tilde{\gamma}_{zz} = 1
    + \frac{\tilde{\gamma}_{yy}\tilde{\gamma}_{xz}^2
    -2\tilde{\gamma}_{xy}\tilde{\gamma}_{yz}\tilde{\gamma}_{xz}
    +\tilde{\gamma}_{xx}\tilde{\gamma}_{yz}^2 }{
    \tilde{\gamma}_{xx}\tilde{\gamma}_{yy}
    -\tilde{\gamma}_{xy}^2}.\label{gzz}
\end{eqnarray}
In all our simulations, we use these two identities instead of
rule AB1.

Combining (\ref{azz}) and (\ref{gzz}) with rule AB2 leads to
exponentially growing, unstable modes if no symmetry assumption is
used.  While it is not clear what exactly causes this instability in
the absence of symmetry assumptions, one possible source is the last
term in the evolution equation (\ref{eq:evolGamma2}) for $\tilde \Gamma^i$,
$(2/3) \tilde{\Gamma}^i\beta^j_{~,j}$.  If
$\beta^j_{~,j} > 0$, as for example for single black holes in
Kerr-Schild coordinates, then this term may lead to exponential growth
of any numerical error in $\tilde \Gamma^i$ (compare a similar
discussion in \cite{lpsd02}).  The sign of this term can be reversed
by adding a multiple of the product ${\mathcal G}^i \beta^j_{~,j}$ to
the evolution equation (\ref{eq:evolGamma2}), specifically
\begin{eqnarray}
\partial_t\tilde{\Gamma}^i&=& \hbox{rhs of (\ref{eq:evolGamma2})} - 
    \left(\chi+\frac{2}{3}\right){\mathcal G}^i\beta^j{}_{,j}\nonumber\\
   &=&2\alpha\left(\tilde{\Gamma}^i_{jk}
    \tilde A^{jk}-\frac{2}{3}\tilde{\gamma}^{ij}K_{,j}
    +6\tilde{A}^{ij}\phi_{,j}\right)-2\tilde{A}^{ij}\alpha_{,j}\nonumber\\
   &&+\tilde{\gamma}^{jk}\beta^i{}_{,jk}+
	\frac{1}{3}\tilde{\gamma}^{ij}\beta^k{}_{,jk}+
	\beta^j\tilde{\Gamma}^i{}_{,j}-\tilde{\Gamma}^j\beta^i{}_{,j}\nonumber\\
   &&+\left[\left(\chi+\frac{2}{3}\right)\tilde{\gamma}^{k\ell}
	\tilde{\Gamma}^i{}_{k\ell}-\chi\tilde{\Gamma}^i\right]
    \beta^j{}_{,j}.\label{Gammat2}
\end{eqnarray}
Here $\chi$ is a free parameter, which in principle can even be chosen
dynamically during an evolution calculation.  To suppress exponential
growth of $\tilde \Gamma^i$, $\chi$ should have the same sign as
$\beta^j_{~,j}$.  

\begin{table*}[t]
\caption{Input parameters for selected evolutions. For each evolution we 
list the lapse condition, the shift condition, the symmetry used,
the parameter $\chi$ in equation~(\ref{Gammat2}), the parameter $\kappa_1$ in
equation~(\ref{modg}), the excision shape, the time when the changes in 
all the variables reach the level of round-off error (if achieved),
and the total run time.}
\begin{ruledtabular}
\begin{tabular}{cccccccccc}
Case\footnote{Cases N3 and N4 have smaller domains; case N4 uses a smaller 
grid spacing (see text).}
&$J/M$&Lapse&Shift\footnote{ Ana, analytic; $\Gamma$: Gamma driver 
(Eq.~\ref{gamma_driver}).}
&Symmetry\footnote{ Oct, octant; Equ, equatorial.}&$\chi$\footnote{
AB2, adopting rule AB2 (Eq.~\ref{AB2})
instead of using Eq.~(\ref{Gammat2}).}&$\kappa_1$&excision&Machine&Run
\\
&&&&&&&&accuracy\footnote{For the symbol ``---'' see Fig.~\ref{aeq9s}.}
&time\footnote{ The symbol ``$>$'' means the run is 
terminated at this time but could continue.} \\ \hline
O0&$0$&Ana&Ana&Oct&AB2&0&cube&No&$<100M$\\
O1&$0$&$1+\log$&Ana&Oct&AB2&0&cube&$1200M$&$>2000M$\\
O2&$0$&$1+\log$&$\Gamma$&Oct&AB2&0&cube&$1000M$&$>2000M$\\
O3&$0$&$1+\log$&Ana&Oct&0&0&cube&$1800M$&$>3000M$\\
O4&$0$&$1+\log$&$\Gamma$&Oct&0&0&cube&$1200M$&$>2000M$\\
O5&$0$&$1+\log$&Ana&Oct&0&0&sphere&$2200M$&$>3000M$\\
O6&$0$&$1+\log$&Ana&Oct&1/3&0&cube&$1600M$&$>2000M$\\ \hline
E0&$0$&$1+\log$&Ana&Equ&AB2&0&cube&No&$1900M$\\
E1&$0$&$1+\log$&Ana&Equ&0&0&cube&$1800M$&$>4000M$\\
E2&$0$&$1+\log$&$\Gamma$&Equ&0&0&cube&$1200M$&$>3000M$\\
E3&$0$&$1+\log$&Ana&Equ&1/3&0&cube&$1600M$&$>3000M$\\ 
E4&$0$&$1+\log$&Ana&Equ&1/3&0&sphere&No&$<200M$\\
E5&$0$&$1+\log$&Ana&Equ&3/4&0&sphere&$1400M$&$>2200M$\\ \hline
N1&$0$&$1+\log$&Ana&None&2/3&0&cube&$1500M$&$>3000M$\\
N2&$0$&$1+\log$&$\Gamma$&None&2/3&0&cube&$1000M$&$>2000M$\\
N3&$0$&$1+\log$&$\Gamma$&None&2/3&0&cube&$700M$&$>1000M$\\
N4&$0$&$1+\log$&$\Gamma$&None&2/3&0&cube&$500M$&$>1000M$\\ \hline
A1&$0.7M$&$1+\log$&Ana&Equ&1/3&0&cube&$3000M$&$>4000M$\\
A2&$0.7M$&$1+\log$&$\Gamma$&Equ&1/3&0&cube&$2000M$&$>3500M$\\
A3&$0.9M$&$1+\log$&Ana&Equ&1/3&0&cube&No&$>1000M$\\
A4&$0.9M$&$1+\log$&Ana&Equ&2/3&0&cube&No&$>3200M$\\
A5&$0.9M$&$1+\log$&Ana&Equ&2/3&0.1&cube&---&$>3300M$\\
A6&$0.9M$&$1+\log$&Ana&Equ&2/3&0.2&cube&---&$>3400M$\\
A7&$0.9M$&$1+\log$&$\Gamma$&Equ&2/3&0&cube&---&$>4600M$ \\
A8&$0.9M$&$1+\log$&$\Gamma$&Equ&2/3&0.1&cube&$4600M$&$>5700M$
\end{tabular}
\end{ruledtabular}
\label{runlist}
\end{table*}

For sufficiently slowly rotating black holes ($J/M \lesssim 0.7 M$),
we were able to follow the evolution for thousands of $M$ without
encountering any instabilities (see Section \ref{nr} and Table
\ref{runlist}).   For more rapidly rotating black holes
($J/M \sim 0.9 M$) we experimented with adjustments suggested
by Yoneda and Shinkai \cite{ygsh02}, namely
\begin{equation} \label{modg}
   \frac{d}{dt}\tilde{\gamma}_{ij}=\hbox{rhs of (\ref{eq:evolg})}
    -\kappa_1 \alpha {\mathcal H} \tilde{\gamma}_{ij}
\end{equation}
and
\begin{equation} \label{moda}
   \frac{d}{dt}\tilde{A}_{ij}=\hbox{rhs of (\ref{eq:evolA})}
    -\kappa_2 \alpha e^{-4\phi}\tilde{\gamma}_{ij} {\mathcal G}^k{}_{,k},
\end{equation}
where $\kappa_1$ and $\kappa_2$ are positive numbers.  We did find
improvements with the adjustment (\ref{modg}), but not with
(\ref{moda}) (see Table \ref{runlist}).  

\section{Numerical Implementation}
\label{ni}

Our numerical implementation follows very closely the recipe
suggested by Alcubierre and Br\"ugmann \cite{ambb01}.  We 
finite-difference the evolution equations using an iterative
Crank-Nicholson scheme with two corrector steps \cite{ts00}.

We use centered differencing everywhere except for the advection
terms on the shift (terms involving $\beta^i\partial_i$).  For
these terms, a second-order upwind scheme is used along the
shift direction.

We adopt ``1+log'' slicing \cite{bcea95,aaea99,amea00}
\begin{equation} \label{1+log}
\partial_t\alpha=D_i\beta^i-\alpha K
\end{equation}
to specify the lapse $\alpha$.  The shift $\beta^i$ is determined
either from the analytic solution, or from the ``Gamma-driver''
condition
\begin{equation} \label{gamma_driver}
\partial_t \beta^i = \lambda\partial_t\tilde{\Gamma}^i
\end{equation}
(see \cite{ambb01}), where we choose $\lambda = 0.05$ in our 
simulations \cite{fn1}.

On the outer boundaries of the numerical grid we impose a radiative
boundary condition that is imposed on the difference between a given
variable and its analytic value $f-f_{\rm analytic}=u(r-t)/r$ where
$u$ is an outgoing wave function.  We apply this condition to all
fields except $\tilde{\Gamma}^i$ which we leave fixed to their
analytic values at the boundary.

We experiment with both cubic and spherical excision regions.  For
cubical excision regions, we adopt the recipe suggested by
\cite{ambb01} to copy the time derivative of every field at the
boundary from its value on a neighboring grid-point.  For surfaces, we
copy from the nearest grid-point along the outward normal, and for
edges and corners from the nearest grid-point along the corresponding
diagonal.

A generalization of this copying algorithm for spherical excision
methods has been suggested in \cite{ybs01}.  For a boundary grid-point
$(i,j,k)$ of the excised sphere we take its nearest neighbors along
the coordinate axes away from the center of the BH, say $(i+1,j,k)$,
$(i,j+1,k)$ and $(i,j,k+1)$.  These three points define a plane, and
we interpolate to the intersection of this plane with the normal on
the surface of the excised region.  If one of these three neighbors is
also a boundary grid-point, we project the normal into the line
defined by the remaining two neighbor points, and do the interpolation
there; if two of the neighbors are inside the excised region we
directly copy the remaining third point.  If all three points are
inside the excised region we average the three diagonal points
$(i+1,j+1,k)$, $(i+1,j,k+1)$ and $(i,j+1,k+1)$; if that is also not
successful we finally copy the point $(i+1,j+1,k+1)$.

We empirically find improved stability if copying along the $x$ or $y$
direction is given higher priority than copying along the $z$
direction.  This asymmetry is probably introduced by (\ref{azz}) and
(\ref{gzz}).  We therefore use the grid-point $(i,j,k+1)$
($(i+1,j,k+1)$ and/or $(i,j+1,k+1)$) only when the grid-points
$(i+1,j,k)$ and $(i,j+1,k)$ ($(i+1,j+1,k)$) are not available.

\begin{figure}[tb]
      \centering
      \leavevmode
      \epsfxsize=3.3in
      \epsffile{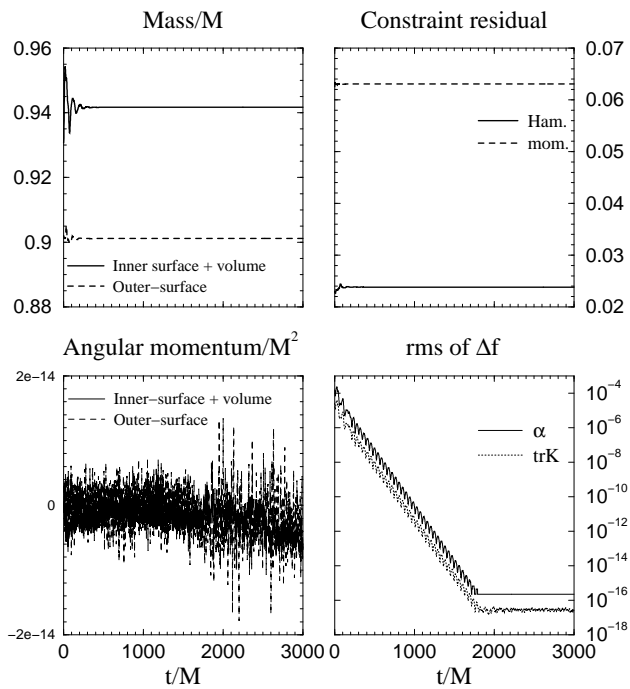}
\caption{The monitored quantities as functions of time for Case O3.
The upper-left panel compares different integrals for the ADM mass.  The
solid line is obtained by using Eq.~(\ref{Mivp}); the dashed line is
obtained by using rhs of Eq.~(\ref{Mout}).  The radius of the inner
surface is $1.5M$ and the radius of the outer surface is $11.5M$.  The
lower-left panel compares different integrals for the angular momentum.
The solid line is obtained by using Eq.~(\ref{Aivp}); the dashed line
is obtained by using rhs of Eq.~(\ref{Aout}).  The upper-right panel
shows the L2 norms of the Hamiltonian constraint ${\mathcal H}$ (the
solid line) and the momentum constraint ${\mathcal M}^x$ (the dashed
line).  The lower-right panel shows a log plot of the root mean square
(r.m.s.) of the changes in the lapse (the solid line) and the trace of
extrinsic curvature (the dotted line) between consecutive time steps.}
\label{oct1}
\end{figure}
\section{Numerical Results}
\label{nr}

Our simulations are summarized in Table \ref{runlist}.  For most of
these simulations we use computational domains of size
$0<x$,$y$,$z<12M$ for octant symmetry, $-12M<x$,$y<12M$ and $0<z<12M$
for equatorial symmetry, and $-12M<x$,$y$,$z<12M$ for no symmetry,
with a grid spacing of $\Delta x=0.4M$.  In order to analyze the
effect of resolution we also performed the two Cases N3 and N4 on a
smaller domain of $-6M<x$,$y$,$z<6M$ and used a resolution of $\Delta
x=0.4M$ for N3 and $\Delta x=0.2M$ for N4.  An additional simulation
with a resolution of $\Delta x=0.8M$ (not included in Table
\ref{runlist}) was used to establish second order convergence of our
code in regions not influenced by the excision surface.  We always use
a Courant factor of $1/4$ so that $\Delta t=\Delta x/4$.  We excise
cubes of volume $(1M)^3$ in octant symmetry, $(2M)^2\times1M$ in
equatorial symmetry, and $(2M)^3$ without symmetry assumptions, or
spheres of radius $1M$.

We will call a simulation ``stable'' if changes in dynamical variables
$\phi$, $\tilde \gamma_{ij}$, $K$, $\tilde A_{ij}$, $\tilde \Gamma^i$,
$\alpha$ and $\beta^i$ drop to round-off error (of about $10^{-16}$ in
double-precision), and remain at that level for several hundred $M$.
Reaching round-off implies that the numerical solution has settled
down to the equilibrium solution of the finite-difference equations
(as opposed to the equilibrium solution of the differential equations,
which is provided as initial data).  In all our stable runs, besides
monitoring the global quantities consisting of the ADM mass, the angular
momentum $J_z$, the $L_2$ norms of the Hamiltonian constraint
$\mathcal H$, the momentum constraint ${\mathcal M}^x$, and the Gamma
constraint ${\mathcal G}^x$ violations, we also monitor the changes in
the representative variables, i.e., $\phi$, $\alpha$, $K$,
$\tilde{\gamma}_{xx}$, and $\tilde{A}_{xx}$, until they reach and
remain at the level of round-off error. Once the equilibrium solution
with a fixed resolution is achieved, we never observe any
instability growing up from the round-off error at later time.

In Table \ref{runlist} we tabulate the time at which any run reaches
equilibrium, and the time after which the simulation is terminated.

Simulations of non-rotating black holes are labeled ``O'' for 
octant symmetry, ``E'' for equatorial symmetry, and ``N'' for
no symmetry.

\begin{figure}[tb]
      \centering
      \leavevmode
      \epsfxsize=3.3in
      \epsffile{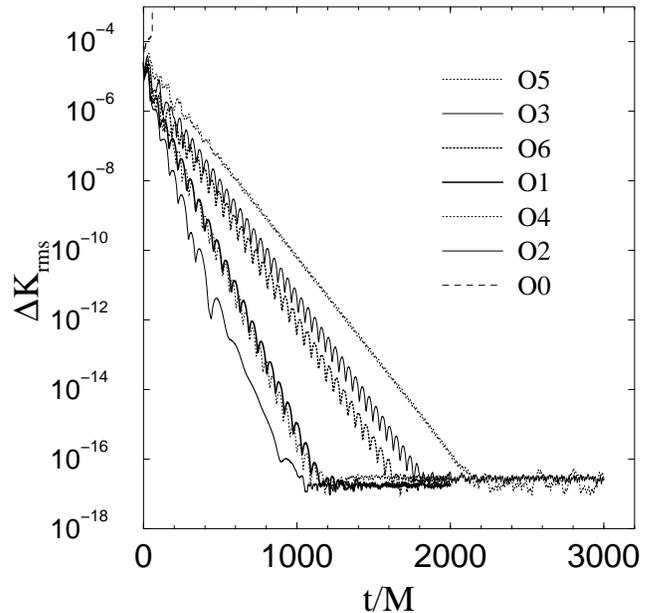}
\caption{The r.m.s. of the change
in the trace of extrinsic curvature between consecutive time steps as
functions of time in the static cases with octant symmetry.  The lines
are labeled sequentially from right to left.  
Case O0 uses a analytic lapse condition.
Recipe AB2 is used in
Cases O1 and O2, while the modification (\ref{Gammat2}) is adopted in
Cases O3 --- O6.  In Case O5 an excision sphere instead of a cube is
used.  All of these cases are stable and the $K$'s all reach their
equilibrium values except Case O0.  
Cases O3 and Case O6 are identical except that
$\chi = 0$ was used in O3 and $\chi = 1/3$ in O6, illustrating 
the effect of the modification (\ref{Gammat2}) in stability.}
\label{oct2}
\end{figure}
\begin{figure}[tb]
      \centering
      \leavevmode
      \epsfxsize=3.3in
      \epsffile{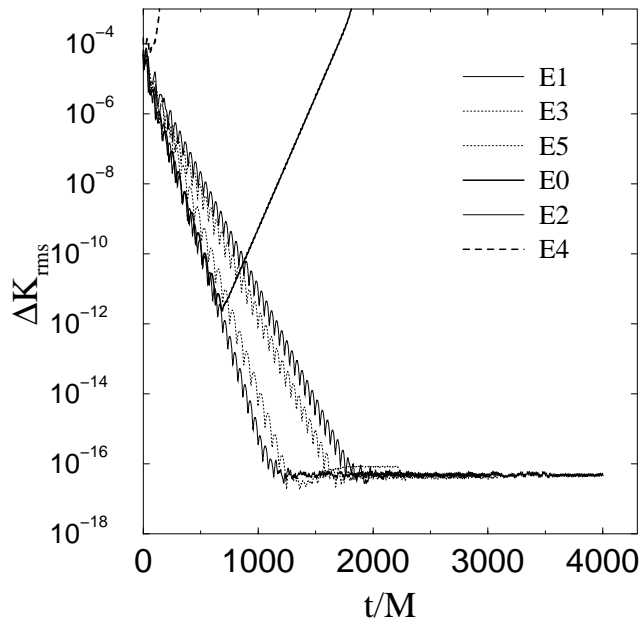}
\caption{The the r.m.s. of the change in the trace of 
extrinsic curvature between consecutive time steps as functions of time 
in the static cases with equatorial symmetry.
The lines are labeled sequentially from right to left.
The recipe AB2 is used in Cases E0.
The modification (\ref{Gammat2}) is used in Cases E1 --- E5 
with different values of $\chi$.
In Case E4 and E5 an excision sphere instead of a cube is used.
All of these cases are stable except Case E0 and E4. 
In E0 an instability appears at $t\sim700M$ and 
the code crashes at $t\sim1900M$.
In E4 the code becomes unstable at the beginning and crashes at $t\sim140M$.
}
\label{equ0}
\end{figure}
\begin{figure}[tb]
      \centering
      \leavevmode
      \epsfxsize=3.3in
      \epsffile{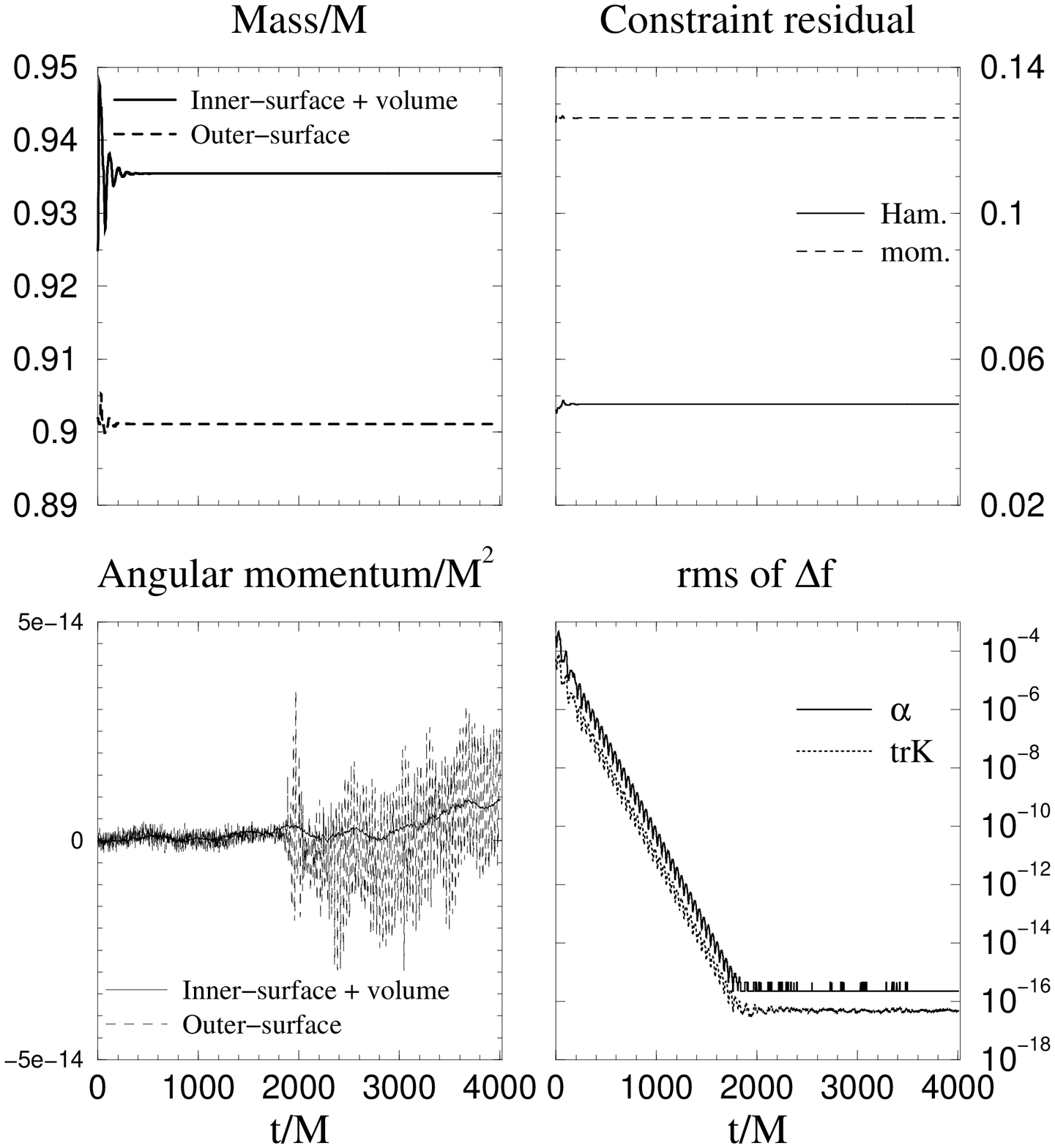}
\caption{The monitored quantities as functions of time for Case E1.
Labeling is the same as in Fig.~\ref{oct1}.}
\label{equ1}
\end{figure}
\begin{figure}[tb]
      \centering
      \leavevmode
      \epsfxsize=3.3in
      \epsffile{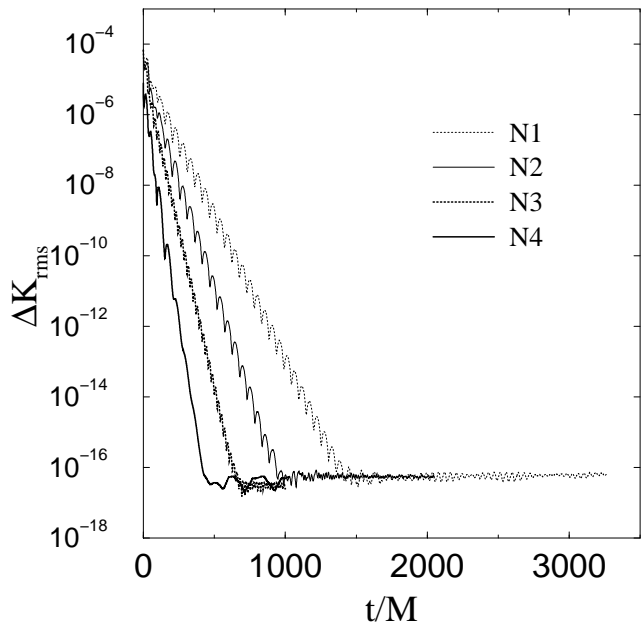}
\caption{The the r.m.s. of the change in the trace of 
extrinsic curvature between consecutive time steps as functions of time 
in the static cases with none symmetry.
The lines are labeled sequentially from right to left.
The settings of case N2 and N3 are the same except the computational
domain of case N3 is half size of case N2 in length. 
The change of $K$ of case N3 drops faster than in 
$K$ of case N2 because that light cross time in case N3 is shorter
than in case N2.
The settings of Case N3 and N4 are the same except the resolution in 
case N4 is two times higher than in case N3.
}
\label{xyz1}
\end{figure}
\begin{figure}[tb]
      \centering
      \leavevmode
      \epsfxsize=3.3in
      \epsffile{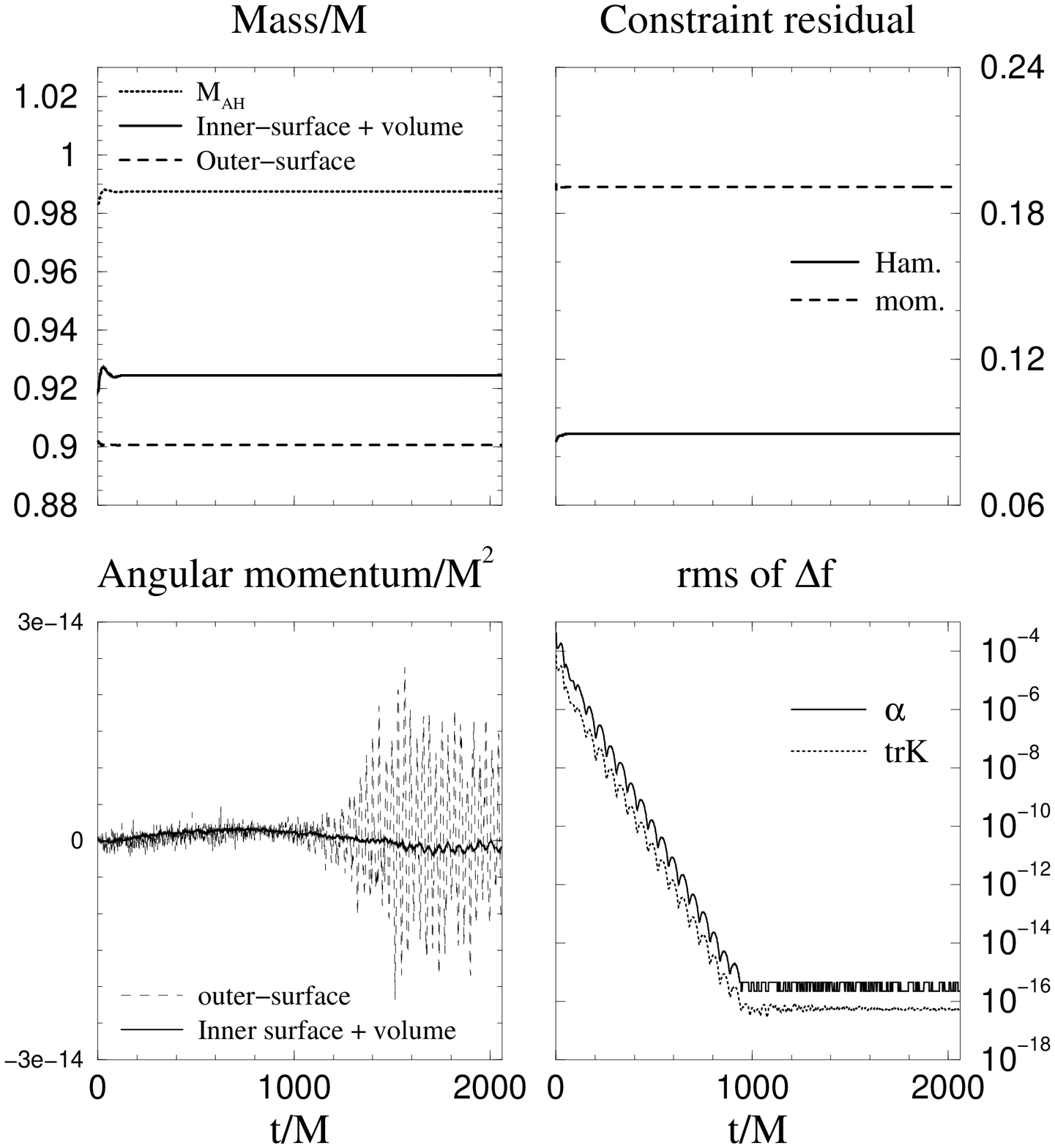}
\caption{The monitored quantities as functions of time for Case N2.
Labeling is the same as in Fig.~\ref{oct1}, except in the upper-left panel
the dotted line is the apparent horizon mass.}
\label{xyz2}
\end{figure}

\subsection{Non-rotating Black Holes}

We first compare several evolutions in octant symmetry.  Cases O0 and
O1 are identical except that in O0 the analytical lapse is used, while
in O1 ``1+log'' slicing (\ref{1+log}) is used.  O0 crashes after a
short time, while O1 is stable.

Cases O1 and O2 confirm that Alcubierre and Br\"ugmann's
\cite{ambb01} algorithm (including their rule AB2) leads to stable
evolution in octant symmetry, both for analytic shift and the
``Gamma-driver'' condition.  The latter consistently allows the
numerical solution to reach equilibrium in a shorter time than the
former.  Simulations O3 --- O6 show that similar results can be
obtained by replacing AB2 with the inclusion of the constraint in
equation (\ref{Gammat2}).  In Case O5 an excision sphere instead of
a cube is used.

Results for case O3 are presented in Fig.~\ref{oct1}.  The upper-left
panel shows two different integrations of the ADM masses which are
derived in Appendix~\ref{ami} and illustrated in Fig.~\ref{fig1}.  The
dashed line is computed from a surface integral at large separation
(equation (\ref{Mout})), while the solid line is computed from a
volume integral plus a surface integral over a small sphere enclosing
the black hole singularity (equation \ref{Mivp}).  We choose a radius
of $R_1 = 1.5 M$ for the inner surface and $R_2 = 11.5 M$ for the
outer surface.  For $R_2 \rightarrow \infty$ the two mass integrals
should agree and should yield the analytic value $M$ of the initial
data.  Our two mass integrals agree to within about 5 \%.  Their
difference arises both from the finite grid spacing, which generates
greater error for surface integrals vs.~volume integrals, as well as
spurious effects near the outer boundary, which may more seriously
affect the outer surface integral.  Their deviation from unity is a
measure of error induced by the proximity of the outer boundary to the
black hole.

The lower-left panel in Fig.~\ref{oct1} shows surface and volume
integrations of the angular momentum, similar to the mass integrations
explained above (see Appendix~\ref{aami}).  The dashed line is
computed from the outer surface integral (\ref{Aout}); the solid line
is computed from a combination of volume integral and inner surface
integral (see equation~(\ref{Aivp})).  For both integrations the 
angular momentum is very close to zero, as it is supposed to be.

The upper-right panel shows the L2 norms of the Hamiltonian constraint
${\mathcal H}$ (solid line) and the momentum constraint ${\mathcal
M}^x$ (dashed line).  The lower-right panel shows a log plot of the
root mean square (r.m.s.) of the changes in the lapse $\alpha$ (the
solid line) and the trace of extrinsic curvature $K$ (the dashed line)
between consecutive time steps.  The changes in $\alpha$ and $K$ both
decrease as exponentially damped oscillations until they reach
round-off error at about $t \sim 1800 M$.  Using the dynamical shift
condition (\ref{gamma_driver}) expedites the damping of these changes
and helps to stabilize the code (compare, for example, cases O1 and O2
or cases O3 and O4).  In Fig.~\ref{oct2} we compare the r.m.s.~of
changes in $K$ for all cases in octant symmetry.

Reconfirming the findings of \cite{ambb01}, we were unable to obtain
stable evolutions with method AB2 if the symmetry is relaxed from
octant to equatorial.  The result of such an evolution (E0) is
included in Fig.~\ref{equ0}, where we plot the r.m.s.~changes of $K$
for various different cases in equatorial symmetry.  For the case E0,
the changes again drop exponentially until $t \sim 700 M$, but at
later times they increase exponentially.  This exponentially growing
mode can be extrapolated back to about round-off error at $t=0$,
indicating that the mode is triggered by round-off error in the
initial data.  Using the Gamma-driver shift condition instead of the
analytic shift leads to similarly unstable evolutions.  However,
replacing the method AB2 with our modification (\ref{Gammat2}) with
$\chi\ge0$, we recover stable evolutions, in which all changes drop
exponentially until they reach round-off error.

In Fig.~\ref{equ0} we compare several different cases in equatorial
symmetry, analyzing the effect of the shift condition, the value of
$\chi$ in (\ref{Gammat2}), and the shape of the excised region.  As
before, we find that using the Gamma-condition instead of the analytic
shift leads to a more rapid settling down to equilibrium, and
comparing cases E1 and E3 shows that increasing $\chi$ has a similar
effect.  Comparing cases E3 and E4, we find that changing the excision
shape from cubic to spherical with the other settings unchanged can
destabilize the code, indicating that our copying method on an excised
sphere leads to larger numerical error than on an excised cube.
Stability can be restored by increasing the value of $\chi$, as for
case E5.  Details for the case E1 are presented in Fig.~\ref{equ1}.

We find that larger values of $\chi$ are needed to obtain long-term
stable evolutions if symmetry assumptions are completely removed.  In
all cases with no symmetry $\chi=2/3$ is used.  We compare the results
of these runs in Figure \ref{xyz1}.  The cases N2 and N3 are identical
except that that the computational domain of N3 is half as large as
that of N2.  We find that N3 settles down to equilibrium faster than
N2, which can be understood in terms of the shorter light crossing
time.  Similarly, cases N3 and N4 are identical except that N4 has
twice the resolution of N3, leading to smaller errors and again faster
approach to equilibrium.

Fig.~\ref{xyz2} shows the result of case N2.  In this evolution we
also locate the apparent horizon every 20 time-steps using the 3D
finder described in \cite{btea96}.  The apparent horizon mass $M_{\rm
AH}= (A/16\pi)^{1/2}$, computed from its area $A$, is included in the
upper-left panel in Fig.~\ref{xyz2} and agrees with the analytic value
to within less than 2 \%.

\begin{figure}[tb]
      \centering
      \leavevmode
      \epsfxsize=3.3in
      \epsffile{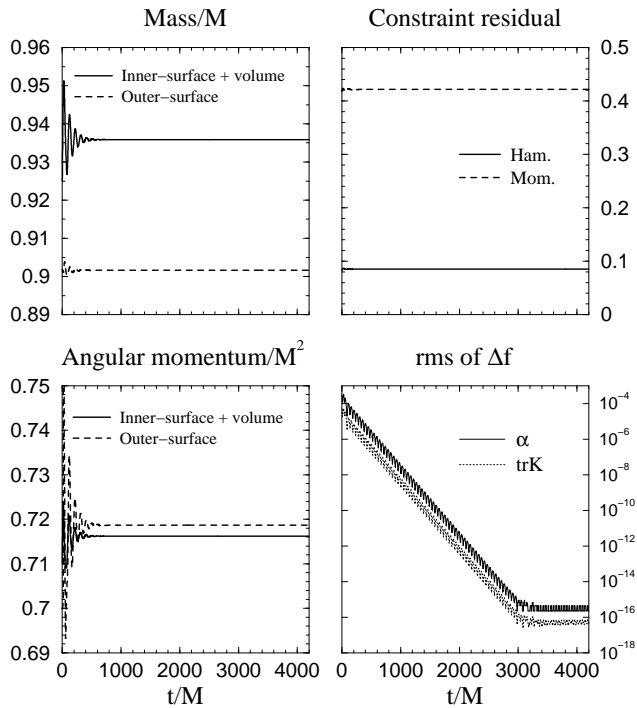}
\caption{The monitored quantities as functions of time for Case A1.
Labeling is the same as in Fig.~\ref{oct1}.}
\label{aeq7}
\end{figure}
\begin{figure}[tb]
      \centering
      \leavevmode
      \epsfxsize=3.3in
      \epsffile{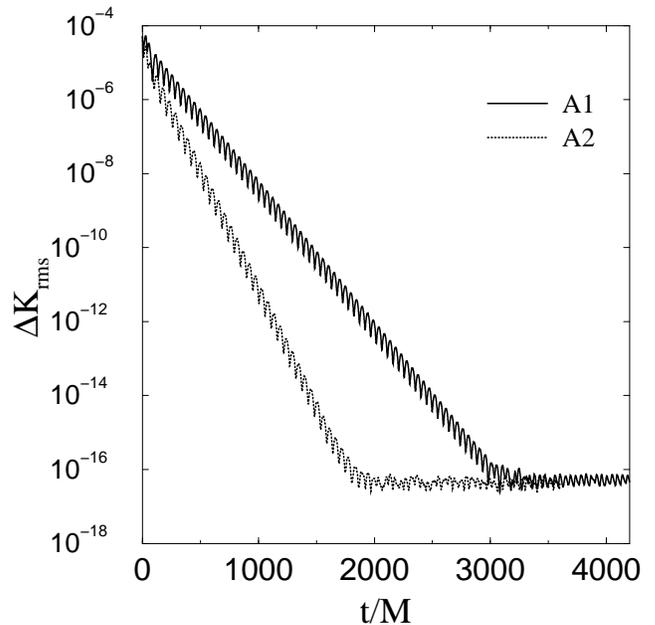}
\caption{The the r.m.s. of the change in the trace of 
extrinsic curvature between consecutive time steps as functions of time 
in the rotating cases with $a=0.7M$.}
\label{aeq7s}
\end{figure}
\begin{figure}[tb]
      \centering
      \leavevmode
      \epsfxsize=3.3in
      \epsffile{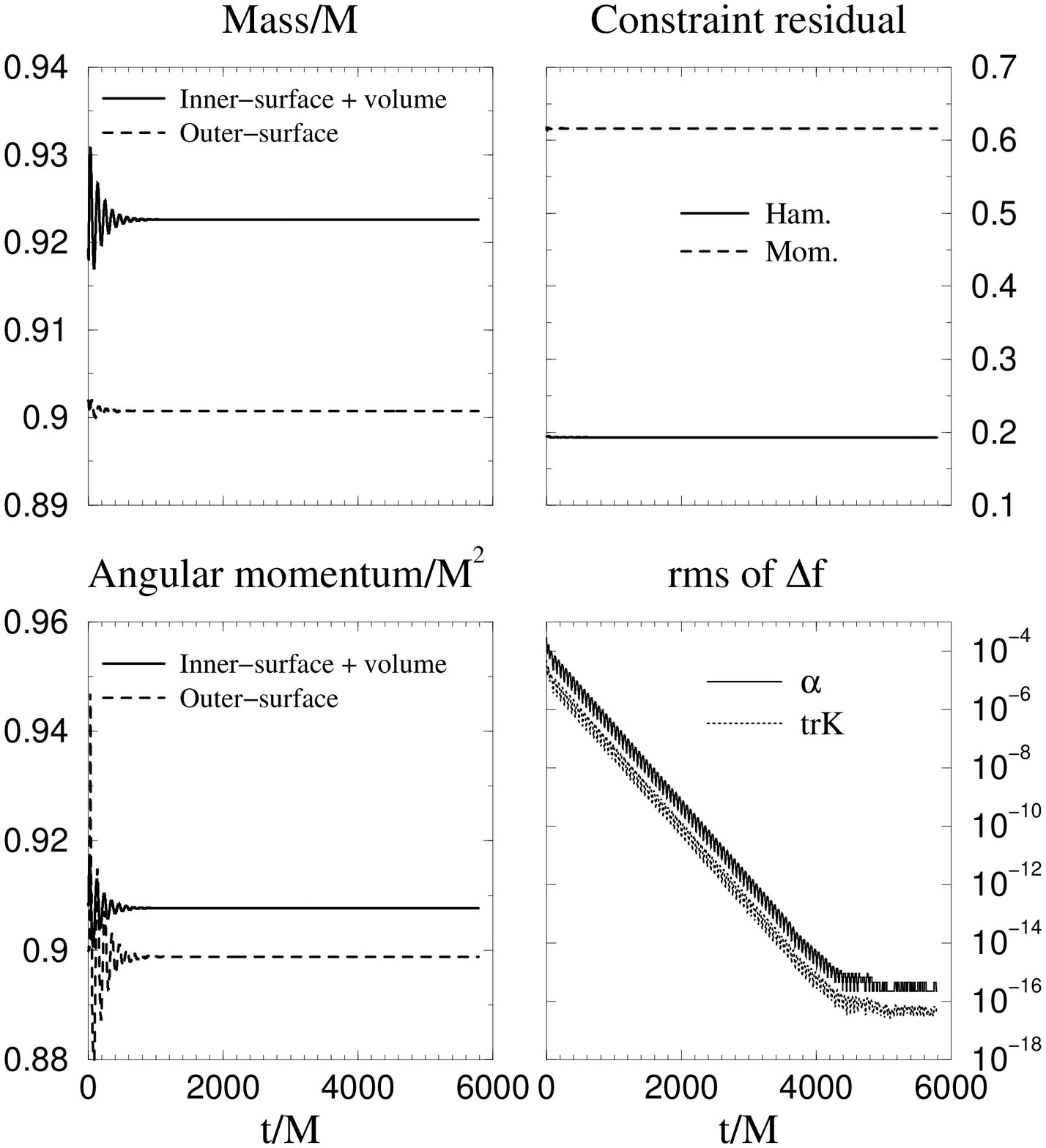}
\caption{The monitored quantities as
functions of time for Case A7.  Labeling is the same as in
Fig.~\ref{oct1}.}
\label{aeq9}
\end{figure}
\begin{figure}[tb]
      \centering
      \leavevmode
      \epsfxsize=3.3in
      \epsffile{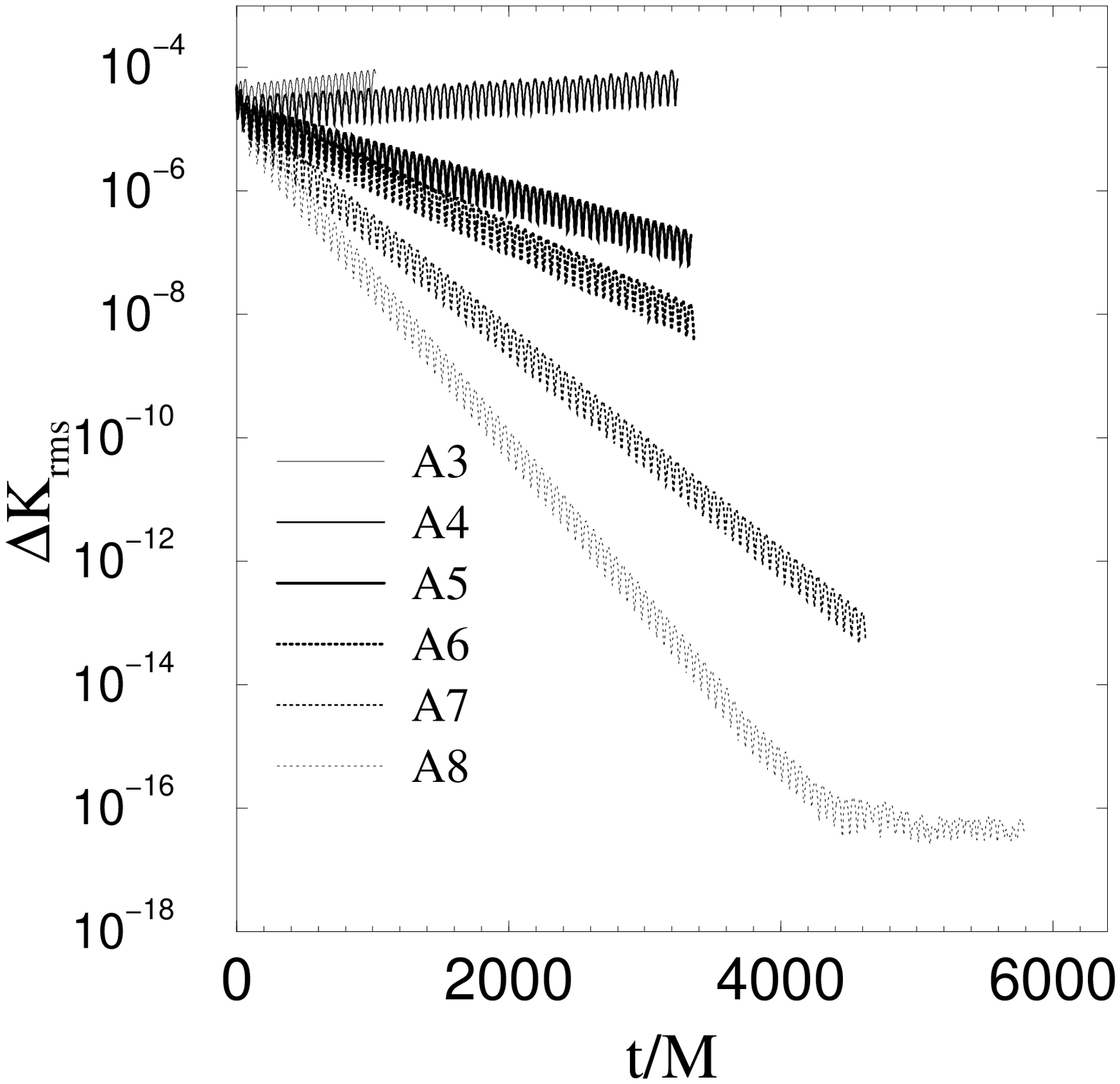}
\caption{The the r.m.s. of the change in the trace of extrinsic
curvature between consecutive time steps as functions of time in the
rotating cases with $a=0.9M$.  The lines are labeled sequentially from
up to down.  The gradual rise of the $\Delta K_{\rm rms}$ in Case A3 and A4
indicates an instability.  The modification (\ref{modg}) is used and
different values of $\kappa$ are tested for stability in Cases A5 and
A6.  The Gamma-driver shift condition is used in both Cases A7 and A8,
and the modification (\ref{modg}) is used in Case A8 with
$\kappa=0.1$.  The improvement of stability can be seen in this figure
by using these modifications.  }
\label{aeq9s}
\end{figure}

\subsection{Rotating Black Holes}

We now turn to black holes with non-zero angular momentum.  Figure
\ref{aeq7} shows the results of Case A1 with a moderate value of the
specific angular momentum $a\equiv J/M=0.7M$.  Comparing with our
results for non-rotating black holes we find that the constraint
violations are larger, and that it takes a longer time for the
solution to settle down to equilibrium ($t\sim3000M$ for A1).  The
mass integrals have larger oscillations at early times than those for
non-rotating black holes, but settle down to similar values as before.
We also find that the values for the angular momentum settle down to
values within about 2 \% of the analytically correct ones.  In
Fig.~\ref{aeq7s} we compare the r.m.s.~of changes in $K$ for A1 and
A2, which again demonstrates that the Gamma-driver shift condition
helps to stabilize the evolution.

Instabilities become even harder to control for $a = 0.9 M$.  The
larger angular momentum leads to larger numerical error, which by
itself makes the simulations more demanding.  In addition, the event
horizon for more rapidly rotating black holes is smaller.  For
$a=0.7M$, our excision cube of physical side length $2M$
(corresponding to a volume of $(2M)^2\times1M$ in equatorial
symmetry), just barely fits inside the event horizon.  For $a=0.9M$,
we decreased the size of the excision cube to $(1.6M)^2\times0.8M$ so
that it does not protrude from the event horizon.  However, we were
unable to obtain stable evolution with the reduced size of the excised
region, which is probably caused by the increasingly large gradients
of the gravitational fields close to the black hole's central
singularity.  Interestingly, we were able to obtain stable evolution
when we left the excision surface at its original size of
$(2M)^2\times1M$.  This surface protrudes from the horizon by small
amounts at the corners of the cube, which introduces errors into the
solution.  However, these errors occur very close to the horizon and
hardly affect the solution in the asymptotic region at all.

In Fig.~\ref{aeq9s} we compare the r.m.s.~of the change in $K$ for
different parameter settings.  Case A3, in which the setting is the
same as case A1 except increasing $a$ from $0.7M$ to $0.9M$, shows an
exponentially growing mode, indicating an instability.  The unstable
mode still cannot be suppressed in case A4 in which$\chi$ has been
increased from $1/3$ to $2/3$.  We find that this instability can be
controlled by adding the Hamiltonian to the evolution equation for the
spatial metric (\ref{modg}) as suggested by \cite{ygsh02} (cases A5
and A6).  Changing from analytical shift to the Gamma-driver shift
condition turns out to be more effective (case A7), and, not
surprisingly, we find the fastest decay of changes in $K$ by combining
both methods (case A8).  We show details of case A8 in
Fig.~\ref{aeq9}.

\section{Summary}
\label{conc}

We experiment with various modifications of the BSSN formulation and
study their effect on the stability of numerical evolution
calculations of static and rotating black holes.  We force the
determinant of the conformally related metric to be unity and the
trace of the traceless part of the extrinsic curvature to be zero.  We
modify the evolution equation for the new auxiliary conformal
connection functions by adding their constraint equation, and also
experiment with adjustments of the other evolution equations suggested
by \cite{ygsh02}.

Most importantly, we find that an instability that arises when octant
symmetry is relaxed \cite{ambb01,s00,klst01,lpsd02,ls02} can be
overcome when the above modifications are employed.  We demonstrate
that both static and moderately rapidly rotating black holes can
be evolved stably without encountering any growing modes.  Any changes
in grid functions settle down to round-off error and remain there for
several 1000 $M$.  

We find that the dynamically enforced ``Gamma-driver'' spatial gauge
condition for the shift leads to more stable evolution than using the
analytical shift.  We also find that cubical excision surfaces, which
are more straight-forward to implement in Cartesian coordinates, work
better than spherical excision surfaces.

While our modifications to not solve all stability problems (e.g.~for
the most extreme rapidly rotating black holes), we believe that they
lead to significant improvements that may be a helpful step towards
simulations of binary black holes and their coalescence.

\acknowledgments

It is a pleasure to thank Matthew Duez and Pedro Marronetti for
helpful discussions.  This work was supported in part by NSF Grants
PHY 00-90310 and PHY-0205155, and NASA Grant NAG 5-10781 at the
University of Illinois at Urbana-Champaign (UIUC), NSF Grant PHY
0139907 at Bowdoin College, and the National Science Council of the
R.O.C. (Taiwan) under grant No.~NSC90007P.  Some of the calculations
were performed at the National Center for Supercomputing Applications
at UIUC.  HJY acknowledges the support of the Academia Sinica, Taipei,
Taiwan.

\appendix
\section{Evaluation of the ADM mass and angular momentum}
\subsection{ADM Mass Integration}
\label{ami}
\begin{figure}
      \centering
      \leavevmode
      \epsfxsize=3.3in
      \epsffile{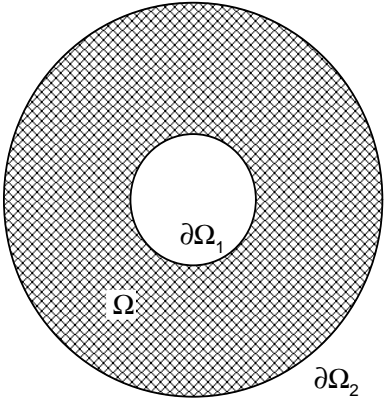}
\caption{The diagram illustrates the relation between the volume
integral on the volume $\Omega$ and the surface integrals on the
boundaries $\partial\Omega_1$ and $\partial\Omega_2$.}
\label{fig1}
\end{figure}

The ADM mass is defined in terms of a surface integral at spatial
infinity.  In numerical simulations, this integral can be approximated
by an integral evaluated on a surface near the outer boundaries of the
grid ($\partial\Omega_2$ in Fig.~\ref{fig1}).  To avoid spurious
effects of the outer boundaries, it is often desirable to convert this
surface integral into a volume integral.  For black hole spacetimes
there is the additional complication of singularities in the black
hole interiors.  In this Appendix, we show how a region inside the
grid can be excluded, so that the mass can be computed from a volume
integral over the outer region and a surface integral over the
boundary of excised interior regions.

In Cartesian coordinates, the ADM mass is defined by a surface
integral at spatial infinity \cite{mnyj74}
\begin{equation}\label{tote}
   M=\frac{1}{16\pi}\oint_\infty \gamma^{im}\gamma^{jn}
    (\gamma_{mn,j} - \gamma_{jn,m}) dS_i.
\end{equation}
where $dS_i\equiv (1/2) \gamma^{1/2} \epsilon_{ijk}{\rm d}x^j{\rm
d}x^k$ is the surface element and $\epsilon_{ijk}$ the Levi-Civita
alternating symbol.  We now perform a conformal decomposition
\begin{equation}\label{barg}
\gamma_{ij} = \psi^{4} \bar \gamma_{ij}.
\end{equation}
where we distinguish from the decomposition (\ref{confgij}) in order
to allow for conformally related metrics $\bar \gamma_{ij}$ with a
determinant different from unity.  Assuming the asymptotic behavior
\begin{equation}
\psi \sim 1 + O(\frac{1}{r}) \mbox{~~~~when~~~~} r \rightarrow \infty
\end{equation}
and 
\begin{equation}
\bar \gamma_{ij} \sim \delta_{ij} + O(\frac{1}{r}) \mbox{~~~~when~~~~}
r \rightarrow \infty,
\end{equation}
we can rewrite (\ref{tote}) as
\begin{eqnarray}
   M &=& \frac{1}{16\pi}\oint_\infty \psi^{-2}
    \bar{\gamma}^{im}\bar{\gamma}^{jn}\left[\psi^4(\bar{\gamma}_{mn,j}
    -\bar{\gamma}_{jn,m})\right. \nonumber\\
   &&\qquad\qquad\qquad\left.+4\psi^3(\psi_{,j}\bar{\gamma}_{mn} 
    -\psi_{,m}\bar{\gamma}_{jn})\right] d\bar{S}_i\nonumber \\
   &= &  \frac{1}{16\pi}\oint_\infty 
    \bar{\gamma}^{im}\left[\bar{\gamma}^{jn}(\bar{\gamma}_{mn,j}
    -\bar{\gamma}_{jn,m})-8\psi_{,m}\right]d\bar{S}_i\nonumber\\
   &=& \frac{1}{16\pi}\oint_\infty (\bar{\Gamma}^i-\bar{\Gamma}^{ji}
    {}_j)d\bar{S}_i
    - \frac{1}{2\pi}\oint_\infty\bar{D}^i\psi d
    \bar{S}_i.\label{newder}
\end{eqnarray}
Here the conformal surface element is defined as $d \bar{S}_i=(1/2)
\bar \gamma^{(1/2)} \epsilon _{ijk}dx^jdx^k$, we use the abbreviations
$\bar{\Gamma}^i\equiv\bar{\gamma}^{jk} \bar{\Gamma}^i{}_{jk}$ and
$\bar \Gamma^{ij}_{~~k} \equiv \bar \gamma^{jl} \bar \Gamma^i_{lk}$,
and $\bar{D}_i$ is the three-covariant derivative with respect to the
metric $\bar{\gamma}_{ij}$.

Using Gauss' law, the surface integral (\ref{newder}) can be converted
into a volume integral.   For spacetimes containing a black
hole, we can exclude an interior region from the volume integration
and write the integral over an outer surface $\partial \Omega_2$ as a
sum of a volume integral over $\Omega$ and a surface integral over an
inner surface $\partial\Omega_1$, 
\begin{equation}
   \int_{\partial\Omega_2} u^id\bar{S}_i
    = \int_\Omega \partial_i(\sqrt{\bar{\gamma}}u^i)d^3x
    + \int_{\partial\Omega_1} u^id\bar{S}_i
\end{equation}
as illustrated in Fig.~\ref{fig1}.  For spacetimes with more than one
black hole, several interior surfaces can be added on the right hand
side.

Applying Gauss' law to the first term of the right hand side of
equation~(\ref{newder}) yields
\begin{eqnarray}
   &&\oint_{\partial\Omega_2} (\bar{\Gamma}^i-\bar{\Gamma}^{ji}
    {}_j)d\bar{S}_i\nonumber\\
    &&=\oint_{\partial\Omega_1} (\bar{\Gamma}^i-\bar{\Gamma}^{ji}
    {}_j)d\bar{S}_i 
    +\int_{\Omega}\partial_i[\sqrt{\bar{\gamma}}(\bar{\Gamma}^i
    -\bar{\Gamma}^{ji}{}_j)]d^3x\nonumber\\ 
   &&=\oint_{\partial\Omega_1} (\bar{\Gamma}^i-\bar{\Gamma}^{ji}
    {}_j)d\bar{S}_i\nonumber\\
   &&\quad+\int_{\Omega} [\partial_i(\bar{\Gamma}^i-\bar{\Gamma}^{ji}{}_j)
    +\bar{\Gamma}^i{}_{ki}(\bar{\Gamma}^k-\bar{\Gamma}^{jk}{}_j)]
    \sqrt{\bar{\gamma}}d^3x\nonumber\\
   &&=\oint_{\partial\Omega_1} (\bar{\Gamma}^i-\bar{\Gamma}^{ji}
    {}_j)d\bar{S}_i\nonumber\\
   &&\quad+\int_{\Omega} (\bar{R}+\bar{\Gamma}^i\bar{\Gamma}^j
    {}_{ij} - \bar{\Gamma}^{ijk} \bar{\Gamma}_{jik})
    \sqrt{\bar{\gamma}}d^3x,
\end{eqnarray}
Similarly, the second term on the right hand side of
equation~(\ref{newder}) yields
\begin{equation}
   \oint_{\partial\Omega_2} \bar{D}^i\psi d\bar{S}_i
     = \int_\Omega \bar{D}^2\psi\sqrt{\bar{\gamma}} d^3x
       + \oint_{\partial\Omega_1}\bar{D}^i\psi d\bar{S}_i,
\label{volint}
\end{equation}
where $\bar{D}^2\equiv\bar{\gamma}^{ij}\bar{D}_i \bar{D}_j$.
Collecting terms, the ADM mass~(\ref{newder}) can now be written
\begin{eqnarray}
   M&=&\frac{1}{16\pi}\oint_{\partial\Omega_2}(\bar{\Gamma}^i
    -\bar{\Gamma}^{ji}{}_j - 8\bar{D}^i\psi)d \bar{S}_i\nonumber\\
   &=&\frac{1}{16\pi}\int_{\Omega}(\bar{R}+\bar{\Gamma}^i
    \bar{\Gamma}^j{}_{ij}-\bar{\Gamma}^{ijk}\bar{\Gamma}_{jik}
    -8\bar{D}^2\psi)\sqrt{\bar{\gamma}}d^3x\nonumber\\
   &&+\frac{1}{16\pi}\oint_{\partial\Omega_1}(\bar{\Gamma}^i
    -\bar{\Gamma}^{ji}{}_j - 8\bar{D}^i\psi)d
    \bar{S}_i.\label{gadmass}
\end{eqnarray}

We now adopt the decomposition of the BSSN formalism (Section
\ref{Bf}), in which the conformal factor is written as $\psi=e^\phi$
and the conformally related metric $\bar{\gamma}_{ij}=
\tilde{\gamma}_{ij}$ is assumed to have determinant $\tilde{\gamma}=1$
so that $\tilde \Gamma^j_{kj} = 0$.  Together with the Hamiltonian
constraint (\ref{ham2})
\begin{equation}\label{ham}
   \tilde{D}^2e^\phi=\frac{e^\phi}{8}\tilde{R} + \frac{e^{5\phi}}{12}K^2 -  
	\frac{e^{5\phi}}{8}
      \tilde{A}_{ij}\tilde{A}^{ij} -2\pi e^{5\phi}\rho,
\end{equation}
where we have included the mass-energy source $\rho \equiv n_\mu n_\nu
T^{\mu\nu}$ for completeness, the ADM mass~(\ref{gadmass}) becomes
\begin{eqnarray}
   M&=& \frac{1}{16\pi} \oint_{\partial\Omega_2} 
        (\tilde{\Gamma}^i-8\tilde{D}^ie^\phi) d\tilde{S}_i\label{Mout}\\
    &=& \frac{1}{16\pi}\int_\Omega d^3x\bigg[e^{5\phi}
        \bigg(16\pi\rho + \tilde{A}_{ij}\tilde{A}^{ij} 
	-\frac{2}{3}K^2\bigg)\nonumber\\
    &&\qquad\qquad\qquad\qquad -\tilde{\Gamma}^{ijk}\tilde{\Gamma}_{jik}
    + (1-e^\phi)\tilde{R}\bigg]\nonumber\\ 
    &&+ \frac{1}{16\pi} \oint_{\partial\Omega_1} 
    (\tilde{\Gamma}^i-8\tilde{D}^ie^\phi) d\tilde{S}_i,\label{Mivp}
\end{eqnarray}
where $d\tilde{S}_i=(1/2)\epsilon_{ijk}dx^jdx^k$ since $\tilde{\gamma}=1$.

We also note that if the conformally related metric falls off faster
than inversely with $r$
\begin{equation}
   \tilde{\gamma}_{ij} = \delta_{ij} + O(1/r^{1+a}),\qquad a>0
\label{falloff}
\end{equation}
(compare \cite{mnyj74}), the first term in equation~(\ref{newder})
vanishes and the mass integral reduces to 
\begin{eqnarray}
   M&=& - \displaystyle{\frac{1}{2\pi} \oint_{\partial\Omega_2}
        \tilde{D}^ie^\phi d\tilde{S}_i}\nonumber\\
    &=& \frac{1}{16\pi}\displaystyle{\int_\Omega d^3x\left[e^{5\phi}
        \left(16\pi\rho + \tilde{A}_{ij}\tilde{A}^{ij} -\frac{2}{3}K^2\right)
        -e^\phi \tilde{R}\right]} 
        \nonumber\\
    &&\displaystyle{- \frac{1}{2\pi} \oint_{\partial\Omega_1} 
       \tilde{D}^ie^\phi d\tilde{S}_i.}\label{yorkm}
\label{1o3}
\end{eqnarray}

Since the Kerr-Schild metric (\ref{ksf1}) does not satisfy the fall-off
condition (\ref{falloff}), equation~(\ref{Mivp}) has to be used to
evaluate its mass (equation (\ref{yorkm}) would yield the incorrect
result $M/3$ for the static Kerr-Schild metric).  We have also found
empirically that, even for metrics for which (\ref{yorkm}) is appropriate, 
(\ref{Mivp}) yields more accurate values for the mass
than~(\ref{yorkm}) in dynamical evolution calculations \cite{dbms02}.

\subsection{ADM Angular Momentum Integration}
\label{aami}

We define the angular momentum $J^i$ as
\begin{equation}\label{Jdef1}
   J_i=\frac{1}{8\pi} \epsilon_{ij}{}^k \oint_\infty x^j A^l{}_k dS_l.
\end{equation}
(compare \cite{york79,by80}), where the indices of $\epsilon_{ij}{}^k$ 
are raised and lowered with the flat metric $\delta_{ij}$.  Since
the integrand is evaluated at $r \rightarrow \infty$, we can replace
$dS_i=e^{6\phi}d\tilde{S}_i$ and $A^i{}_j=\tilde{A}^i{}_j$ and use
Gauss' law to obtain
\begin{eqnarray}
   8\pi J_i&=&\epsilon_{ij}{}^k \oint_{\partial\Omega_2} x^j
       e^{6\phi}\tilde{A}^l{}_k d\tilde{S}_l\nonumber\\
   &=& \epsilon_{ij}{}^k \oint_{\partial\Omega_1} 
       x^je^{6\phi}\tilde{A}^l{}_k d\tilde{S}_l
   +\epsilon_{ij}{}^k\int_\Omega(x^je^{6\phi}\tilde{A}^l{}_k)_{,l}d^3x.
   \nonumber\\
\label{Jgau}
\end{eqnarray}
The volume integral in Eq.~(\ref{Jgau}) is
\begin{eqnarray}
   &&\int_\Omega(x^je^{6\phi}\tilde{A}^l{}_k)_{,l}d^3x\nonumber\\
   &&=\int_\Omega\left[\delta^j{}_le^{6\phi}\tilde{A}^l{}_k 
     +x^j(e^{6\phi}\tilde{A}^l{}_k)_{,l}\right]d^3x \\
   &&=\int_\Omega\left[e^{6\phi}\tilde{A}^j{}_k
     +x^j\tilde{D}_l(e^{6\phi}\tilde{A}^l{}_k)
     +e^{6\phi}x^j\tilde{\Gamma}^n{}_{kl}\tilde{A}^l{}_n\right]d^3x\nonumber\\
   &&=\int_\Omega\left[e^{6\phi}\tilde{A}^j{}_k
     +x^j\tilde{D}_l(e^{6\phi}\tilde{A}^l{}_k)
     -\frac{1}{2}e^{6\phi}x^j\tilde{A}_{ln}\tilde{\gamma}^{ln}{}_{,k}
     \right]d^3x\nonumber
\label{Jvol}
\end{eqnarray}
where we have used $\tilde \gamma = 1$.  With the momentum constraint,
\begin{equation}
    \tilde{D}_j(e^{6\phi}\tilde{A}^j{}_i)
    =e^{6\phi}\left(\frac{2}{3}\tilde{D}_iK + 8\pi s_i\right),
\end{equation}
where we have included the momentum 
density $s_i\equiv\gamma_{i\mu}n_\nu T^{\mu\nu}$ for completeness,
the volume integral (\ref{Jvol}) can be rewritten
\begin{eqnarray}
   J_i &=&\frac{1}{8\pi} \epsilon_{ij}{}^k \oint_{\partial\Omega_2}
           e^{6\phi}x^j \tilde{A}^l{}_k d\tilde{S}_l\label{Aout}\\
    &=&\frac{1}{8\pi} \epsilon_{ij}{}^k\int_\Omega
           \bigg[e^{6\phi}\bigg(\tilde{A}^j{}_k 
           + \frac{2}{3}x^j\tilde{D}_kK\nonumber\\
    &&\qquad\qquad\quad- \frac{1}{2} x^j\tilde{A}_{ln}
     \partial_k\tilde{\gamma}^{ln}{}+8\pi x^js_k\bigg)\bigg]d^3x\nonumber\\
    &&+\frac{1}{8\pi} \epsilon_{ij}{}^k \oint_{\partial\Omega_1}
           e^{6\phi} x^j \tilde{A}^l{}_k d\tilde{S}_l.\label{Aivp}
\end{eqnarray}
This expression is equivalent to Eq.~(2.25) of \cite{shim2}.  Note
that the first term does not vanish identically, since indices of
$\epsilon_{ijn}$ are raised with $\delta_{ij}$ while indices of
$\tilde{A}^{jn}$ are lowered with $\tilde{\gamma}_{ij}$.



\begin{thebibliography}{99}

\bibitem{adm62} R. Arnowitt, S. Deser and C. W. Misner, in 
        {\em Gravitation: An Introduction to Current Research}, 
        edited by L. Witten (Wiley, New York, 1962).

\bibitem{bonc2}
C. Bona {\it et al}, Phys. Rev. Lett. {\bf 75}, 600 (1995).

\bibitem{baut}
T. W. Baumgarte and S. L. Shapiro, Phys. Rev. D {\bf 59}, 024007 (1999).

\bibitem{ay99} A. Anderson and J. W. York, Jr., 
	Phys. Rev. Lett. {\bf 82}, 4384 (1999).

\bibitem{r98} O. Reula, Living Rev. Rel. {\bf 1}, 3 (1998).

\bibitem{shim}
M. Shibata and T. Nakamura, Phys. Rev. D {\bf 52}, 5428 (1995).

\bibitem{baut2}
T. W. Baumgarte, S. A. Hughes, and S. L. Shapiro, Phys. Rev. D {\bf 60} 087501
(1999).

\bibitem{shim2}
M. Shibata, Phys. Rev. D {\bf 60}, 104052 (1999).

\bibitem{shim3}
M. Shibata and K. Uryu, Phys. Rev. D {\bf 61}, 064001 (2000).

\bibitem{shim4}
M. Shibata, T. W. Baumgarte, and S. L. Shapiro, Phys. Rev. D {\bf 61},
044012 (2000).

\bibitem{alcm4}
M. Alcubierre {\it et al}, Phys. Rev. D {\bf 61}, 041501 (2000).

\bibitem{alcm5}
M. Alcubierre {\it et al}, Phys. Rev. D {\bf 62}, 124011 (2000).

\bibitem{smal}
L. Smarr and J. York, Phys. Rev. D {\bf 17}, 2529 (1978).

\bibitem{eard}
D. Eardley and L. Smarr, Phys. Rev. D {\bf 19}, 2239 (1979).

\bibitem{barj}
J. M. Bardeen and T. Piran, Phys. Rep. {\bf 196}, 205 (1983).

\bibitem{plst8586}
L.I. Petrich, S.L. Shapiro, and S.A. Teukolsky, Phys. Rev. D {\bf 31},
2459 (1985); {\bf 33}, 2100 (1986).

\bibitem{unrw84}
W.G. Unruh, as cited in \cite{thoj2} (1984).

\bibitem{thoj2}
J. Thornburg, Class. Quantum Grav. {\bf 4}, 1119 (1987).

\bibitem{seie}
E. Seidel and W.-M. Suen, Phys. Rev. Lett. {\bf 69}, 1845 (1992).

\bibitem{thoj3}
J. Thornburg, PhD. thesis, Univ. of British Columbia, Vancouver, British
Columbia, 1993.

\bibitem{annp}
P. Anninos {\it et al}, Phys. Rev. D {\bf 51}, 5562 (1995).

\bibitem{schm2}
M. A. Scheel, S. L. Shapiro and S. A. Teukolsky, Phys. Rev. D {\bf 51}, 4208
(1995).

\bibitem{bras}
S. Brandt {\it et al}, Phys. Rev. Lett. {\bf 85}, 5496 (2000).

\bibitem{mrhs99}
R. A. Matzner, M. F. Huq, and D. Shoemaker, 
Phys. Rev. D {\bf 59}, 024015 (1999).

\bibitem{bisn}
N. T. Bishop {\it et al}, Phys. Rev. D {\bf 57}, 6113 (1998).

\bibitem{marp}
P. Marronetti and R. A. Matzner, Phys. Rev. Lett. {\bf 85}, 5500 (2000).

\bibitem{stp02}
O. Sarbarch, M. Tiglio, and J. Pullin, Phys. Rev. D {\bf 65}, 064026 (2002).

\bibitem{pct02}
H. P. Pfeiffer, G. B. Cook, and S. A. Teukolsky, gr-qc/0203085 (2002).

\bibitem{ambb01}
M. Alcubierre and B. Br\"ugmann, Phys. Rev. D {\bf 63}, 104006 (2001).

\bibitem{s00} M. A. Scheel, talk given at the ITP Miniprogram ``Colliding
	Black Holes: Mathematical Issues in Numerical Relativity'',
	January 10 -- 28, 2000.

\bibitem{lpsd02}
P. Laguna and D. Shoemaker, gr-qc/0202105 (2002).

\bibitem{ls02}
L. Lindblom and M. Scheel, gr-qc/0206035 (2002).

\bibitem{klst01}
L. E. Kidder, M. A. Scheel, and S. A. Teukolsky, Phys. Rev. D {\bf 64}, 064017
(2001).

\bibitem{d87}
S. Detweiler, Phys. Rev. D {\bf 35}, 1095 (1987).

\bibitem{ys01} 
G. Yoneda and H. Shinkai, Phys. Rev. D {\bf 63}, 124019 (2001).

\bibitem{kllpsst01}
B. Kelly, P. Laguna, K. Lockitch, J. Pullin, E. Schnetter, 
D. Shoemaker, and M. Tiglio, 
Phys. Rev. D {\bf 64}, 084013 (2001).

\bibitem{ygsh02}
G. Yoneda and H. Shinkai, gr-qc/0204002.

\bibitem{kwb02}  A. M. Knapp, E. J. Walker, T. W. Baumgarte, 
	 Phys. Rev. D {\bf 65}, 064031 (2002).

\bibitem{chas92}
S. Chandrasekhar, {\it The Mathematical Theory of Black Holes} (Oxford
University Press, New York, 1992).

\bibitem{edda58}
A. E. Eddington, Nature {\bf 113}, 192 (1924); D. Finkelstein, Phys. Rev.
{\bf 110}, 965 (1958).

\bibitem{ts00}
S. Teukolsky, Phys. Rev. D {\bf 61}, 087501 (2000).

\bibitem{bcea95}
C. Bona {\it et al.}, Phys. Rev. Lett. {\bf 75}, 600 (1995).

\bibitem{aaea99}
A. Arbona {\it et al.}, Phys. Rev. D {\bf 60}, 104014 (1999).

\bibitem{amea00}
M. Alcubierre {\it et al.}, Phys. Rev. D {\bf 62}, 044034 (2000).

\bibitem{fn1} The Gamma-driver condition (equations (\ref{Gammat2})
and (\ref{gamma_driver})) is a parabolic equation for the shift,
$\partial_t\beta^i\sim\lambda\tilde{\gamma}^{jk}\beta^i{}_{jk}$.
Accordingly, stability requires that we set $\lambda\propto(\Delta
x)^2/\Delta t\propto\Delta x$ when $\Delta x/\Delta t$ is fixed by a
hyperbolic Courant condition.

\bibitem{ybs01}
H.-J. Yo, T. W. Baumgarte, and S. L. Shapiro, 
Phys. Rev. D {\bf 64}, 124011 (2001).

\bibitem{btea96}
T. W. Baumgarte, {\it et al.}, Phys. Rev. D {\bf 54}, 4849 (1996).

\bibitem{mnyj74} 
N. \'O Murchadha and J. W. York, Jr., Phys. Rev. D {\bf 10}, 2345, (1974).

\bibitem{dbms02} M. D. Duez, T. W. Baumgarte, P. Marronetti, and
S. L. Shapiro, in preparation, 2002.

\bibitem{york79} J. W. York, Jr., in {\it Sources of Gravitational
Radiation}, edited by L.L. Smarr (Cambridge Univ. Press, Cambridge,
1979).

\bibitem{by80} J. M. Bowen and J. W. York, Jr., Phys. Rev. D {\bf 21}
2047 (1980).


\end{thebibliography}
\end{document}